\documentclass[twocolumn, prl,superscriptaddress,preprintnumbers,showpacs,amsmath,amssymb,floatfix]{revtex4}

\usepackage{graphicx}
\usepackage{dcolumn}

\usepackage{bm}

\begin{document}

\title{Tailoring the flow of soft glasses by soft additives}

\author{E.~Zaccarelli}
\affiliation{Dipartimento di Fisica and CNR-INFM-SOFT,
Universit{\`a} di Roma La Sapienza,  I-00185 Rome, Italy}
\affiliation{ISC-CNR, Via dei Taurini 19, I-00185 Rome, Italy}

\author{C.~Mayer}
\affiliation{Institut f{\"u}r Theoretische Physik II,
             Heinrich-Heine-Universit{\"a}t,
             D-40225 D{\"u}sseldorf,
             Germany}

\author{A.~Asteriadi}
\affiliation{FO.R.T.H., Institute of Electronic Structure and Laser,
GR-71110 Heraklion, Crete, Greece}

\author{C.~N.~Likos}
\affiliation{Institut f{\"u}r Theoretische Physik II,
             Heinrich-Heine-Universit{\"a}t,
             D-40225 D{\"u}sseldorf,             
             Germany}

\author{F.~Sciortino} 
\affiliation{Dipartimento di Fisica and CNR-INFM-SOFT,
Universit{\`a} di Roma La Sapienza,  I-00185 Rome, Italy}

\author{J.~Roovers}
\affiliation{NRC, Institute for Chemical Process and Environmental
Technology, Ottawa, Ontario, Canada}

\author{H.~Iatrou}
\affiliation{University of Athens, Department of Chemistry, GR-15771
Athens,
Greece}

\author{N.~Hadjichristidis}
\affiliation{University of Athens, Department of Chemistry, GR-15771
Athens,
Greece}
\author{P.~Tartaglia}
\affiliation{Dipartimento di Fisica and CNR-INFM-SMC, 
Universit{\`a} di Roma La Sapienza, I-00185 Rome, Italy}

\author{H.~L{\"o}wen}
\affiliation{Institut f{\"u}r Theoretische Physik II,
             Heinrich-Heine-Universit{\"a}t,
             D-40225 D{\"u}sseldorf,
             Germany}

\author{D.~Vlassopoulos}
\affiliation{FO.R.T.H., Institute of Electronic Structure and Laser,
GR-71110 Heraklion, Crete, Greece}
\affiliation{University of Crete, Department of Materials Science
and Technology, GR-71003 Heraklion, Crete, Greece}

\begin{abstract}
We examine the vitrification and melting of asymmetric star polymers mixtures
by combining rheological measurements with mode coupling theory.
We identify two types of glassy states, a {\it single} glass, 
in which the small component is fluid in the glassy matrix of the big 
one and a {\it double} glass, in which both components are vitrified. 
Addition of small star polymers leads to melting 
of {\it both} glasses and the melting curve has
a non-monotonic dependence on the star-star size ratio. 
The phenomenon opens new ways for externally steering 
the rheological behavior of soft matter systems.
\end{abstract}


\date{\today} 

\maketitle

The design of materials with well-defined rheological properties
and the ability to alter these 
at wish, by tuning
suitable control parameters of the system,
are issues of central importance in today's soft matter research.
Experimental findings, accumulating at a fast pace, call for the
identification and profound understanding of the several underlying
physical mechanisms that control the ability of
soft materials to support stresses or flow under shear \cite{cates:book}.
In several situations, e.g., in coating or processing applications,
it is necessary to dramatically alter the viscoelastic properties
of a material. One possibility to do so in a controlled way,
is to exploit the phenomenon of dynamical arrest. Indeed, close to a
liquid-glass transition, a small variation of external parameters
produces a spectacular change in the elastic properties of the material
without significantly affecting its structure \cite{dimitris2004}.

Star polymers have emerged as 
an ideal model system for exploring the flow behavior of soft matter
and elucidating its molecular origin.
They consist of $f$ polymer chains covalently anchored onto a common
center \cite{Grest:review:96}. Star polymer solutions
are chemically simple, well-characterized 
and physically tunable in their softness, 
building thereby natural bridges between hard colloids
and polymers. This property stems from the particular
form of the effective, entropic interaction between star-polymer 
centers \cite{Witten:Pincus,likos},  which has a logarithmic/Yukawa 
form at small/large separations.
Technologically, star polymers are important in several 
applications such as their use as viscosity
modifiers in the oil industry \cite{Grest:review:96} or their
novel applications as 
drug-delivery agents \cite{djordjevic:etal:03, yang:03}.

In this work, we show how one can gain control over the 
the rheological properties of soft, repulsive, glassy materials via the
addition of a second,
soft, repulsive component, which is chemically identical to the first
but smaller in size.
We consider
a star-polymer glass perturbed by smaller
star-polymer additives. We observe melting of the
big-star glass, induced by the small stars, and an unexpected non-monotonic
dependence of the critical amount of additives,
needed to melt the glass, on the big-small star size ratio.
We also perform the corresponding
calculations of the melting line, within the binary mode coupling
theory (MCT) framework \cite{goetze,goetze2}.
The theoretical melting line is in agreement with 
the experimental one, reproducing the qualitative 
trends of the latter. We demonstrate that the
non-monotonic behavior arises from
two different mechanisms by which
the presence of the additive significantly affects the rheological
properties, depending on the size ratios between the components.
The former rests on the fluidity of the smaller component in
the {\it single glass} formed by the larger component. The latter
stems from the mutual soft repulsions in the {\it double glass},
in which both components are vitrified. The melting of such
a double glass
represents a novel physical process, in which 
a glassy component is liquified 
through the addition of a second glassy component.

\begin{figure}
\begin{center}
\includegraphics[width=5.5cm,clip=true]{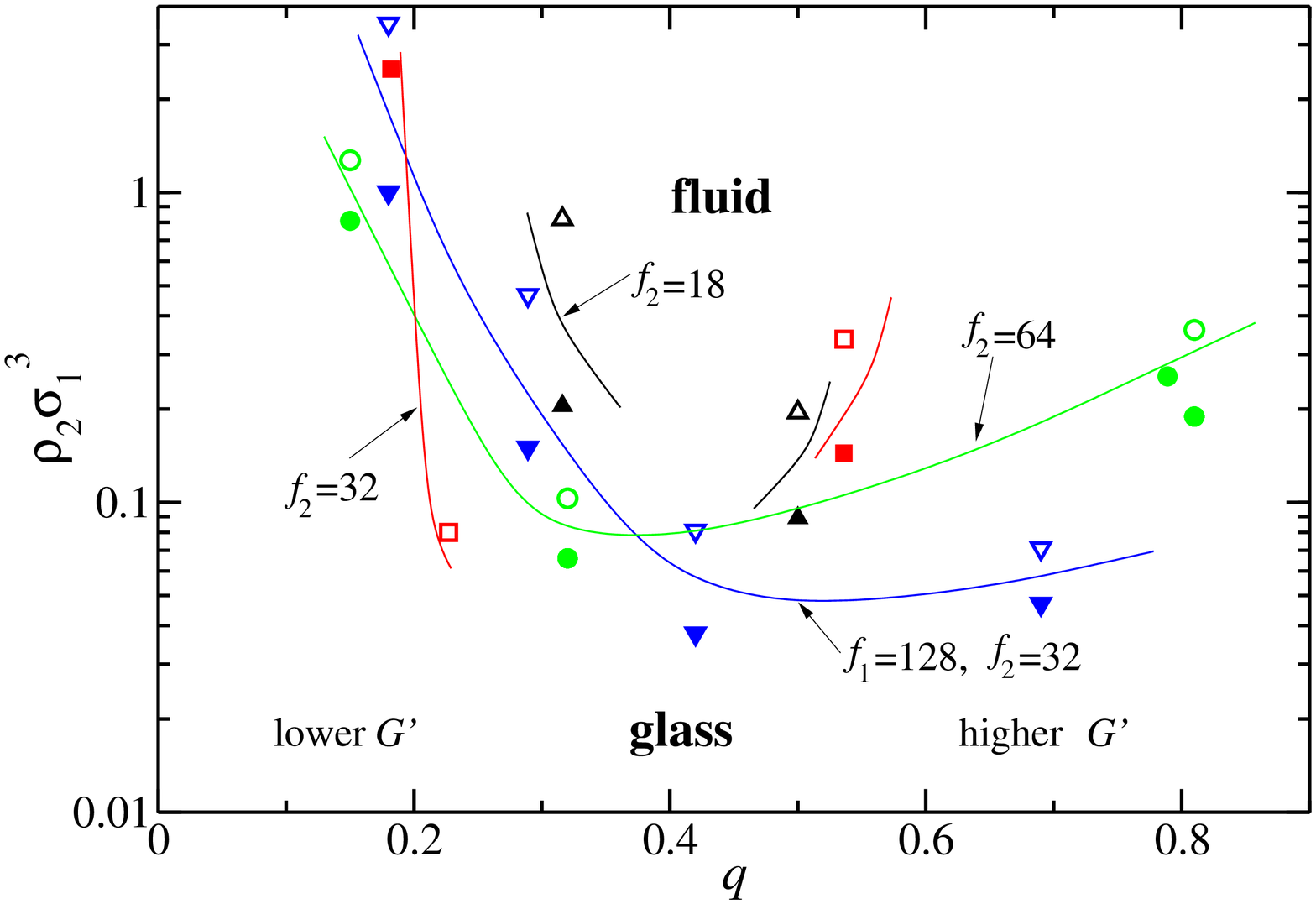}
\end{center}
\caption{Experimental kinetic `phase diagram' of binary star mixtures,
indicating regions of liquid and glass for different concentrations of
added small star $\rho_2$ and size ratios $q$. Large stars 
of functionalities $f_1 = 270$ and $128$ were used, with densities
respectively $\rho_1\sigma_1^3 = 0.345$ and $0.412$.  Open symbols
denote a liquid and closed ones a solid (glass) state. 
Only the
data points closest to the melting lines are shown here.
The lines through the data, passing above the full symbols and below
the empty ones, are guides to the eye.}
\end{figure}

Binary mixtures of big and small (1,4-polybutadiene)-stars were
prepared in toluene. Three different types of big stars were employed,
having an average functionality $f_1 \cong 270$ arms and arm molecular
weight $M_{\rm a}$ ranging from $18\,000$ to $42\,000\,{\rm g/mol}$
\cite{roovers:89}. Soft glasses were obtained at concentration
$c_1/c_1^{*} \cong 1.4$, where $c_1^{*}$ denotes the big-star overlap
concentration. Additional measurements were also performed with
big stars of functionality $f_1=128$ with arm molecular weight of
about 80000 g/mol, which are very similar to the
$270$-stars \cite{jpcm2001}. We express the mixture composition by
the values $\rho_1\sigma_1^3$ of the big and $\rho_2\sigma_1^3$ of the
small stars, where $\rho_i$, $i=1,2$ are the respective number
densities and $\sigma_i$ are the corona diameters of the big stars
appearing in the effective interactions employed in the theory,
coinciding with the stars' hydrodynamic radii $R_{h,i}$, $i=1,2$.
Small stars with three different functionalities, $f_2 = 16$, $32$,
and $64$, and molecular weights $M_{\rm a}$ between $1200$ and
$80\,000\,{\rm g/mol}$ \cite{toporowski:86, zhou:92, roovers:93}, were
synthesized as well.  Size ratios $q \equiv R_{h,2}/R_{h,1} =
\sigma_2/\sigma_1$ varied from $0.15$ to $1$.  The mixture preparation
protocol consisted of creating the big star glass at fixed number
density $\rho_1\sigma_1^3 = 0.345$ ($c_1/c_1^{*} = 1.4$) and then
adding small stars with certain $q$ at a desired density $\rho_2$,
under conditions of very gentle and prolonged stirring.
In this procedure, the glass was broken and the
mixture was left to `equilibrate' again.  For the 
$f_1=128$-sample the fixed density was
$\rho_1\sigma_1^3 =0.412$.  Dynamical rheological measurements (time,
strain and frequency sweeps) were carried out in order to identify the
state of the particular samples (solid or liquid behaviour).  A
strain-controlled rheometer was utilized in the cone-and-plate
geometry ($25\,{\rm mm}$ diameter, $0.04\,{\rm rad}$ cone angle) and
dynamic frequency sweep tests were conducted in the range $100 -
0.1\,{\rm rad/s}$ at $20\,^{\circ}{\rm C}$ in the linear viscoelastic
regime.

The experiments carried out provide evidence of U-shaped melting
curves in the plane spanned by the size ratio $q$ and the number
density $\rho_2$ of the small additives, as well as quantitative
distinctions of two types of glasses depending on the value of $q$.
The glassy samples are characterized by the typical virtually
frequency-independent elastic moduli $G'$ ($350\,{\rm Pa}
<G'<800\,{\rm Pa}$) and respective weakly-frequency dependent viscous
moduli $G''$ that exhibit a broad minimum ($G''_{\rm min}$$20\,{\rm
Pa}<G''_{\rm min}< 55\,{\rm Pa}$) \cite{mason:95}.  For small
concentration of small stars, the system exhibits.  moduli $G' > G''$
and $G' \sim \omega^{0}$, characteristic of solid, glass-like behavior
\cite{mason:95, larson:99, raynaud:96}.  As the density of the added
small star increases, there is a dramatic change in the mixture's
viscoelastic response. The glass melts, and a fluid results, whose
moduli are weaker by orders of magnitude.  A compilation of
rheological results obtained from the available samples of different
$q$ and $\rho_2$ is presented in Fig.\ 1 for both studied
functionalities of the large stars.  In order to have 
a clear representation of all results in
one plot, only the data closest to the glass-liquid boundaries are
shown. This figure represents kinetic `phase diagrams' of binary
star mixtures, at fixed $f_1$ and $\rho_1$ and varying small-star
characteristics in the $(q,\rho_2)$-parameter space.  The results
suggest that a U-shaped melting line separates the fluid (above) from
the glass (below) states. This trend is
general, since it is
independent of the big stars
used in the experiment.

Our theoretical analysis is based on a coarse-grained star description 
employing effective interactions \cite{likos} in binary star
mixtures \cite{ferber} 
with functionalities $f_i$ and corona diameters $\sigma_i$,
$i = 1,2$, combined with mode coupling
theory (MCT) \cite{goetze,goetze2}. In the pure star solution,
MCT predicts
a glass line at large $f$ above an $f$-dependent
density \cite{prl}, in agreement
with experiment \cite{dimitris2004}.
The equilibrium structure factors $S_{ij}(k)$, $k = 1, 2$,
were calculated by solving the
two-component Ornstein-Zernike equation \cite{hansen} within the
Rogers-Young closure \cite{ry}. 
The dynamics 
was calculated using two-component MCT \cite{goetze3, barrat}.
 
Upon addition of the smaller stars, the partial static structure
factor $S_{11}(k)$ of the large ones shows a loss of structure, a phenomenon
caused by the depletion-induced softening of the repulsions between
the big stars. At the same time, the small-stars partial structure
factor, $S_{22}(k)$, gradually develops growing peaks. These
structural changes result into melting of the glass at a $q$- and
$f_2$-dependent density $\rho_{2}^{\rm melt}$. Our 
theoretical results are summarized in Fig.\ 2, which can be
interpreted as the theoretical `kinetic phase diagram' of the system, in 
analogy with its experimental counterpart of Fig.\ 1. In
agreement with experimental results, a characteristic U-shape is found
for the melting curves at all studied $f_2$. 
Above $q=0.4$ for $f_2=64$, and $q=0.45$ for $f_2=16$
and $32$, no melting is found for any density of the small stars.

\begin{figure}
\begin{center}
\includegraphics[width=5.5cm,clip=true]{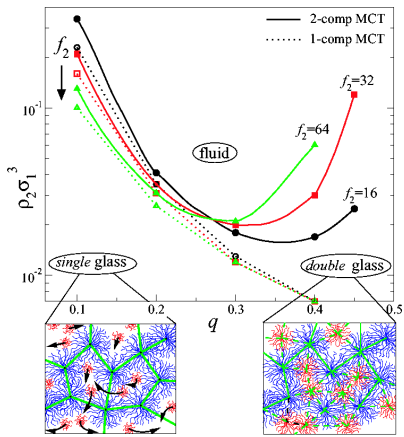}
\end{center}
\caption{ Theoretical kinetic `phase diagram' of binary star mixtures,
calculated using MCT.  The large-star functionality and concentration
are fixed at the values $f_1 = 270$ and $\rho_1\sigma_1^3 = 0.345$.
The diagram is shown
for three different
functionalities $f_2$ of the small stars. Circles: $f_2 = 16$; squares: $f_2 =
32$; triangles: $f_2 = 64$. The lines going through the calculated
points are guides to the eye. The cartoons display local arrangements
in a single big-star glass with mobile small stars (left) and in a
double glass in which there is mutual caging of both components (right).}  
\end{figure}

The U-shape of the melting line points to the existence of two
distinct microscopic melting mechanisms.  For very small size of the
additives, a standard depletion mechanism takes place. The osmotic
pressure from the mobile, small stars, which are free to diffuse in
the matrix formed by the big ones, leads to a reduction of the
repulsive interactions between the latter. As a consequence, the cages
that stabilize the glass are weakened, eventually breaking at a
sufficiently high small-star concentration, at which the system melts.
Since depletion is stronger at fixed $\rho_2$ as $q$ grows, the
melting curve has, in this regime, a negative slope in the
$(q,\rho_{2})$-plane, generating the left arm of the U-shaped
line. However, as $q$ further increases, the smaller stars become
themselves slower and they begin to actively participate in the glass
formation.  At low additives concentration, the latter become trapped
in the voids left out of the glassy matrix.  Hence, two competing
mechanisms are at work: as before, the structure of the big stars is
weakened by the addition of the small ones but, at the same time, the
second component becomes increasingly glassy. The onset of this
mechanism brings about a reversal of the slope of the melting curve,
since now the tendency of the small stars to soften the repulsion
between the big ones is counter-driven by their own opposing tendency
to jam. As $\rho_2$ grows at fixed $q$, the small-star jamming cannot
persist indefinitely, since the direct repulsions between the
additives prevent them from occupying the same region of free
space. As a result, melting of the glass takes place.  We can
distinguish between two different glassy states under the U-curve, one
in which only the big stars are jammed (low $q$) and one in which both
components are arrested (large $q$).  We call the former {\it
single-glass} and the latter {\it double-glass} state.

The above interpretation is supported, first, by a comparison with the
results obtained if a
one-component 
MCT-treatment is adopted, in which the smaller stars are assumed to
form an ergodic fluid (dotted lines in Fig.\ 2). 
The two
approaches yield very similar results at small size ratios, whereas
the discrepancy becomes large at higher $q$'s, signaling the tendency
of the small component to arrest.  The deviation between the two
approaches can also be understood in terms of the scaling of the
short-time mobilities 
with size ratio
\cite{zacca2004}: as $q \rightarrow 1$ the time-scale separation
between the 
two species
disappears.

\begin{figure}
\begin{center}
\includegraphics[width=5.5cm,clip=true]{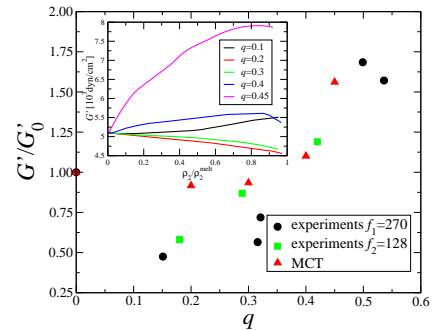}
\end{center}
\caption{ 
Plot of the ratio $G'/G'_0$, where $G'_0$ is the
modulus of the system without additives, against  size ratio $q$
close to melting. Theory refers to the case $f_{2} = 32$, while 
experiments refer to a 
compilation of several measured elastic moduli of star-star mixtures,
obtained with different big and small stars at various concentrations
of the additives.
Inset: Dependence of the
elastic modulus $G'$ of the glass on $\rho_2$ ($f_2 = 32$).}
\end{figure}

Further, we have calculated the elastic modulus $G'$ following 
Ref.\ \cite{naegele:jcp:98}.  In the inset of Fig.\ 3 we show the
theoretical results for $G'$, demonstrating that, for $q \leq 0.3$,
$G'$ has a maximum at $\rho_{2} = 0$, whereas for the larger size
ratio, $q = 0.4$, it has a minimum there. 
At small
$q$, the additives lower $G'$ through the softening of the big cages,
whereas at high $q$ they lead to stiffening of the glass through the
fact that they are themselves driven to dynamical arrest.  In Fig.\ 3,
theoretical and experimental results for the {\it normalized} modulus
$G'/G'_0$ are shown where $G'_0 \sim 500\,{\rm Pa}$ stands for the
average value of the elastic modulus of the big-star glass, without
any additives. Experimental results display the same trends predicted
theoretically. For small $q$, experimental values are much lower
than the theoretical ones
because MCT is not capable of
taking into account the vast discrepancy in the mobilities of the two
species in highly asymmetric mixtures. However, at larger $q$, in a
full binary regime where the theory works best, the agreement between
theory and experiments becomes almost quantitative.

Additional evidence for the fluidity of the small stars at low
$q$-values and their jammed nature at high ones is offered by the
calculated nonergodicity factor $f_{22}(k)$, shown in  
Fig.\ 4.  
Whereas the big-star nonergodicity
factor $f_{11}(k)$ is rather insensitive to $q$, and remains roughly
the same in all glassy states, $f_{22}(k)$ shows a dramatic change.
For $q \leq 0.2$, $f_{22}(k)$ is very small and its nonzero values are
confined to a very narrow, small $k$-domain.  
Such non-ergodicity factors are fully consistent with a mobile small
component \cite{bosse,bosse2,bosse3}. At high $q$-values 
$f_{22}(k)$ has the typical range 
of a fully arrested system.

\begin{figure}[ht]
\begin{center}
\includegraphics[width=5.5cm,clip=true]{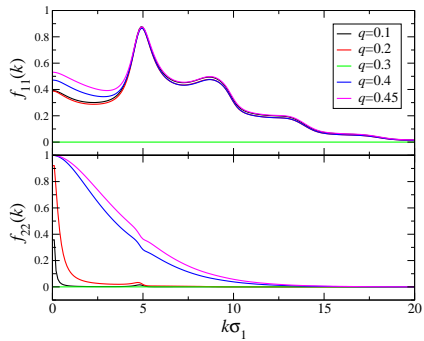}
\end{center}
\caption{
The nonergodicity factors 
$f_{11}(k)$ (top) and
$f_{22}(k)$ (bottom) of the two
components,
for $f_2 = 32$, $\rho_2\sigma_1^3 = 0.02$ at several $q$-values.
For $q = 0.3$ the system is melted.
}
\end{figure}

The phenomenon of glass melting 
through polymer addition 
has been observed in 
hard-colloid--polymer (CP) mixtures \cite{pham, bartsch}.
The physics and implications of this are, 
however, in our case very different. Whereas in the CP-case
the solid melts due to the induction
of short-range attractions on the colloids, which eventually drive the
formation of a reentrant, attractive glass, in our case effective
attractions are completely absent. This fundamental difference is
witnessed by the fact that in CP-mixtures melting takes place only for
very asymmetric size ratios, $q \leq 0.1$, whereas in the present case
it persists up to $q \cong 1$.  The melting of
the solid in CP-mixtures {\it requires} the fluidity of
the added polymer \cite{zacca2004}. To the contrary, for
star mixtures, the addition of
a second component that becomes increasingly {\it glassy} is 
capable of bringing about a melting of the double glass.
This is made possible by the 
crucial role played by the softness of the interactions
between all 
species, which allows for the 
and rearrangement of {\it soft} cages, a mechanism absent in
CP-mixtures. In star-linear mixtures, 
only the left arm of the U-curve is seen experimentally, 
the melting phenomenon ceasing altogether at 
$q \cong 0.5$ [\onlinecite{manolis:prl:02}]. There, the 
linear chains form a transient physical network, 
whereas stars are 
forced 
to maintain their 
spherical shape.

Our findings are general, since they are based on the softness
of the effective interactions involved.  A large class of composite
soft materials, such as charge-stabilized colloidal dispersions,
microgels, dendritic
polymers or self-organized 
micelles are characterized by soft effective interactions. 
Charged suspensions and dusty plasmas,
for instance, can be described by Yukawa potentials
whose decay length and strength can be tuned by controlling the
charges and the screening microions, in a similar way that the
strength and range of the effective interactions between star
polymers can be tuned by changing their functionality. Therefore,
they should be amenable to manipulations of their rheology along the
lines reported here. Functional versatility and viscosity control
are desired material properties in medical, pharmaceutical and
coating applications as well. 
Controlling these products near vitrification transitions,
provides a means to achieve the desired
properties variation.

We thank E.~Stiakakis for assistance with some measurements.
This work has been supported by MIUR FIRB,
MRTN-CT-2003-504712, by the DFG
within the SFB-TR6, and by the EU within the NoE
``Softcomp''. C.M.\ thanks the D{\"u}sseldorf
Entrepreneurs Foundation for a Ph.D.\ Fellowship.

\end{document}